\documentclass[aps,prl,preprint,showpacs]{revtex4} 
\usepackage{graphicx}

\begin{document}

\title{Ginzburg -- Landau expansion in strongly disordered attractive
Anderson -- Hubbard model}

\author{E.Z. Kuchinskii$^1$, N.A. Kuleeva$^1$, M.V. Sadovskii$^1$$^,$$^2$}


\affiliation{$^1$Institute for Electrophysics, Russian Academy of Sciences, Ural
Branch, Ekaterinburg 620016, Russia\\
$^2$M.N. Mikheev Institute for Metal Physics, Russian Academy of Sciences, Ural
Branch,
Ekaterinburg 620990, Russia}


\begin{abstract}

We have studied disordering effects on the coefficients of Ginzburg -- Landau
expansion in powers of superconducting order -- parameter in attractive
Anderson -- Hubbard model within the generalized DMFT+$\Sigma$ approximation.
We consider the wide region of attractive potentials $U$ from the weak coupling
region, where superconductivity is described by BCS model, to the strong coupling
region, where superconducting transition is related with Bose -- Einstein
condensation (BEC) of compact Cooper pairs formed at temperatures essentially
larger than the temperature of superconducting transition, and the wide range
of disorder --- from weak to strong, where the system is in the vicinity
of Anderson transition. In case of semi -- elliptic bare density of states
disorder influence upon the coefficients $A$ and $B$ before the square and the
fourth power of the order -- parameter is universal for any value of electron
correlation and is related only to the general disorder widening of the bare
band (generalized Anderson theorem). Such universality is absent for the
gradient term expansion coefficient $C$. In the usual theory of ``dirty''
superconductors the $C$ coefficient drops with the growth of disorder.
In the limit of strong disorder in BCS limit the coefficient $C$ is very
sensitive to the effects of Anderson localization, which lead to its further
drop with disorder growth up to the region of Anderson insulator.
In the region of BCS -- BEC crossover and in BEC limit the coefficient $C$ and
all related physical properties are weakly dependent on disorder.
In particular, this leads to relatively weak disorder dependence of both
penetration depth and coherence lengths, as well as of related slope of the
upper critical magnetic field at superconducting transition, in the region of 
very strong coupling.
 
\end{abstract}


\maketitle

\section{Introduction}

The studies of disorder influence on superconductivity have rather long history.
The pioneer works by Abrikosov and Gor'kov \cite{AG_impr,AG_imp,Gor_GL,AG_mimp} 
considered the limit of weak disorder ($p_Fl\gg 1$, where $p_F$ is the Fermi
momentum and $l$ is the mean free path) and weak coupling superconductivity
well describe by BCS theory. The notorious ``Anderson theorem'' on
superconducting critical temperature $T_c$ of superconductors with ``normal''
(non magnetic) disorder  \cite{And_th,Genn} is usually also referred to these
limits.

The generalization of the theory of ``dirty'' superconductors to the case of
strong enough disorder ($p_Fl\sim1$) (and further up to the region of Anderson
transition) was made in Refs. \cite{SCLoc_1,SCLoc_2,SCLoc_3}, where
superconductivity was also considered in the weak coupling limit.

The problem of BCS theory generalization to the strong coupling region is studied
also for a long time. The significant progress in this direction was achieved
by Nozieres and Schmitt-Rink \cite{NS}, who proposed an effective method to
study the crossover from BCS -- type behavior in the weak coupling region
to Bose -- Einstein condensation (BEC) in the strong coupling region. At the
same time the problem of superconductivity of disordered systems in the limit
of strong coupling and in BCS -- BEC crossover region remains relatively
undeveloped.

One of the simplest models to study the BCS -- BEC crossover is the attractive
Hubbard model. The most successful approach to the studies of Hubbard model,
both to describe strongly correlated systems in case of repulsive interactions
and to study BCS -- BEC crossover in case of attraction, is the dynamical
mean -- field theory (DMFT) \cite{pruschke,georges96,Vollh10}.

In recent years we have developed the generalized DMFT+$\Sigma$ approach to
Hubbard model \cite{JTL05,PRB05,FNT06,UFN12,HubDis,LVK16}, which is very
convenient to the description of different additional ``external'' (as compared
to DMFT) interactions. In particular, this approach is well suited to describe
also the two -- particle properties, such as optical (dynamic) conductivity
\cite{HubDis,PRB07}.

In Ref. \cite{JETP14} we have used this approach to analyze single -- particle
properties of the normal phase and optical conductivity in the attractive
Hubbard model. Further on, DMFT+$\Sigma$ method was used by us in Ref.
\cite{JTL14} to study disorder effects on superconducting critical temperature,
which was calculated within Nozieres -- Schmitt-Rink approach. In particular,
for the case of semi -- elliptic model of the bare density of states, which is
adequate to describe three -- dimensional systems, we have demonstrated
numerically, that disorder influence upon the critical temperature (for the
whole range of interaction parameters) is related only to the general widening
of the bare band (density of states) by disorder. In Ref. \cite{JETP15} we have
presented an analytic derivation of such disorder influence (in DMFT+$\Sigma$
approximation) on all single -- particle properties and the temperature of
superconducting transition for the case of semi -- elliptic band.

Starting with classic paper by Gor'kov \cite{Gor_GL} ii is well known, that
Ginzburg -- Landau expansion plays the fundamental role in the theory of
``dirty'' superconductors, allowing the effective treatment of disorder
dependence of different physical properties close to superconducting
critical temperature \cite{Genn}. The generalization of this theory to the
region of strong disorder (up to Anderson metal -- insulator transition) was
also based upon microscopic derivation of the coefficients of this expansion
\cite{SCLoc_1,SCLoc_2,SCLoc_3}. However, as noted above, all these derivations
were performed in the weak coupling limit of BCS theory.

In Ref. \cite{JETP16} we have combined the Nozieres -- Schmitt-Rink and
DMFT+$\Sigma$ approximations within the attractive Hubbard model to derive
coefficients of homogeneous Ginzburg -- Landau expansion $A$ and $B$ before the
square and the fourth power of superconducting order -- parameter, demonstrating
the universal disorder influence on coefficients $A$ and $B$ and the related
discontinuity of specific heat at the transition temperature. After that, in
Ref. \cite{FNT16} we have studied the behavior of coefficient $C$ before the
gradient term of Ginzburg -- Landau expansion, where such universality is absent.
In this work we have only considered this coefficient in the region of weak
disorder ($p_Fl\gg 1$) in  the ``ladder'' approximation for impurity scattering,
as it is usually done in the standard theory of ``dirty'' superconductors
\cite{Gor_GL}, though for the whole range of pairing interactions including the
BCS -- BEC crossover region and the limit of very strong coupling. In fact, here
we have neglected the effects of Anderson localization, which can significantly
change the behavior of the coefficient $C$  in the limit of strong disorder
($p_Fl\sim 1$) \cite{SCLoc_1,SCLoc_2,SCLoc_3}.

In the current work we shall concentrate mainly on the study of the coefficient
$C$ in the region of strong disorder, when Anderson localization effects become
relevant.

\section{Hubbard model within DMFT+$\Sigma$ approach and Nozieres -- Schmitt-Rink
approximation}

We consider the disordered nonmagnetic attractive Anderson -- Hubbard model,
described by the Hamiltonian:
\begin{equation}
H=-t\sum_{\langle ij\rangle \sigma }a_{i\sigma }^{\dagger }a_{j\sigma
}+\sum_{i\sigma }\epsilon _{i}n_{i\sigma }-U\sum_{i}n_{i\uparrow
}n_{i\downarrow },  
\label{And_Hubb}
\end{equation}
where $t>0$ is transfer amplitude between nearest neighbors, $U$ is the
Hubbard -- like onsite attraction,
$n_{i\sigma }=a_{i\sigma }^{\dagger }a_{i\sigma }^{{\phantom{\dagger}}}$ is
electron number operator at a given site,
$a_{i\sigma }$ ($a_{i\sigma }^{\dagger}$) is annihilation (creation) operator
of an electron with spin $\sigma$, and local energies $\epsilon _{i}$ are
assumed to be independent random variables at different lattice sites.
For the validity of the standard ``impurity'' diagram technique \cite{Diagr,AGD}
we assume the Gaussian distribution for energy levels $\epsilon _{i}$:
\begin{equation}
\mathcal{P}(\epsilon _{i})=\frac{1}{\sqrt{2\pi}W}\exp\left
(-\frac{\epsilon_{i}^2}{2W^2}
\right)
\label{Gauss}
\end{equation}
Distribution width $W$ is the measure of disorder, while the Gaussian field of
energy levels (independent on different sites -- ``white'' noise correlation)
induces the ``impurity'' scattering, which is described by the standard
approach, based upon the calculation of the averaged Green's functions
\cite{Diagr}.

The generalized DMFT+$\Sigma$ approach \cite{JTL05,PRB05,FNT06,UFN12} extends
the standard dynamical mean -- field theory (DMFT)
\cite{pruschke,georges96,Vollh10} introducing the additional ``external''
self -- energy part (SEP) $\Sigma_{\bf p}(\varepsilon)$
(in general momentum dependent), which originates from any interaction outside
the DMFT, and provides an effective procedure to calculate both singe -- particle
and two -- particle properties \cite{HubDis,PRB07}. The success of such
generalized approach is connected with the choice of single -- particle
Green's function in the following form:
\begin{equation}
G(\varepsilon,{\bf p})=\frac{1}{\varepsilon+\mu-\varepsilon({\bf p})-
\Sigma(\varepsilon)
-\Sigma_{\bf p}(\varepsilon)},
\label{Gk}
\end{equation}
where $\varepsilon({\bf p})$ is the ``bare'' electronic dispersion, while the
total SEP is an additive sum of Hubbard -- like local SEP $\Sigma (\varepsilon)$
and ``external'' $\Sigma_{\bf p}(\varepsilon)$, neglecting the interference
between Hubbard -- like and ``external'' interactions. This allows to conserve
the system of self -- consistent equations of the standard DMFT
\cite{pruschke,georges96,Vollh10}. At the each step of DMFT iterations the
the ``external'' SEP $\Sigma_{\bf p}(\varepsilon)$ is recalculated with the use
of some approximate scheme, corresponding to the form of additional interaction,
while the local Green's function is also ``dressed'' by
$\Sigma_{\bf p}(\varepsilon)$ at each step of the standard DMFT procedure.

The ``external'' SEP, entering DMFT+$\Sigma$ cycle, in the problem of disorder
scattering under consideration here \cite{HubDis,LVK16},  is taken in the
simplest (self -- consistent Born) approximation, neglecting the ``crossing''
diagrams of impurity scattering, which gives:
\begin{equation}
\Sigma_{\bf p}(\varepsilon)\to\Sigma_{imp}(\varepsilon)=W^2\sum_{\bf p}
G(\varepsilon,{\bf p}),
\label{BornSigma}
\end{equation}

To solve the effective single Anderson impurity problem of DMFT we use here,
as in our previous papers, the quite efficient impurity solver using the
numerical renormalization group (NRG) \cite{NRGrev}.

In the following we are using the ``bare'' band with semi -- elliptic density
of states (per unit cell with lattice parameter $a$ and single spin projection),
which is rather good approximation in three -- dimensional case:
\begin{equation}
N_0(\varepsilon)=\frac{2}{\pi D^2}\sqrt{D^2-\varepsilon^2}
\label{DOSd3}
\end{equation}
where $D$ defines the half -- width of the conduction band.

In Ref. \cite{JETP15} we have shown that in DMFT+$\Sigma$ approach for the
model with semi -- elliptic density of states all effect of disorder upon
single -- particle properties reduces only to the band -- widening due to
disorder, i.e. to the replacement $D\to D_{eff}$, where $D_{eff}$ is the
effective half -- width of the ``bare'' band in the absence of electronic
correlations($U=0$), widened by disorder:
\begin{equation}
D_{eff}=D\sqrt{1+4\frac{W^2}{D^2}}.
\label{Deff}
\end{equation}
The ``bare'' density of states (in the absence of $U$) ``dressed'' by disorder:
\begin{equation}
\tilde N_{0}(\xi)=\frac{2}{\pi D_{eff}^2}\sqrt{D_{eff}^2-\varepsilon^2}
\label{tildeDOS}
\end{equation}
remains semi -- elliptic also in the presence of disorder.
It should be noted, that in other models of the ``bare'' band disorder effect is
not reduced only to the widening of the band, changing also the form of the
density of states, so that there is no complete universality of disorder
influence on single -- particle properties, reducing to a simple substitution
$D\to D_{eff}$. However, in the limit of strong enough disorder of interest to
us, the ``bare'' band becomes practically semi -- elliptic restoring such
universality \cite{JETP15}.

All calculations below, as in our previous works, were performed for rather
typical case of quarter -- filled band (the number of electrons per lattice
site is n=0.5).

To consider superconductivity for the wide range of pairing interaction $U$,
following Refs. \cite{JETP14,JETP15}, we use Nozieres -- Schmitt-Rink
approximation \cite{NS}, which allows qualitatively correct (though approximate)
description of BCS -- BEC crossover region. In this approach we determine the
critical temperature $T_c$ using the usual BCS -- type equation \cite{JETP15}:
\begin{equation}
1=\frac{U}{2}\int_{-\infty}^{\infty}d\varepsilon \tilde N_0(\varepsilon)
\frac{th\frac{\varepsilon -\mu}{2T_c}}{\varepsilon -\mu},
\label{BCS}
\end{equation}
with chemical potential $\mu$ determined via DMFT+$\Sigma$ calculations for
different values of $U$ and $W$, i.e. from the standard equation for the number
of electrons (band filling), determined by the Green's function given by
Eq. (\ref{Gk}), allowing us to find $T_c$  for the wide range of the model
parameters including the regions of BCS -- BEC crossover and strong coupling,
as well as for different levels of disorder.
This reflects the physical meaning of Nozieres -- Schmitt-Rink approximation ---
in the weak coupling region transition temperature is controlled by the
equation for Cooper instability (\ref{BCS}), while in the strong coupling region
it is determined as BEC temperature controlled by chemical potential.

\begin{figure}
\includegraphics[clip=true,width=0.7\textwidth]{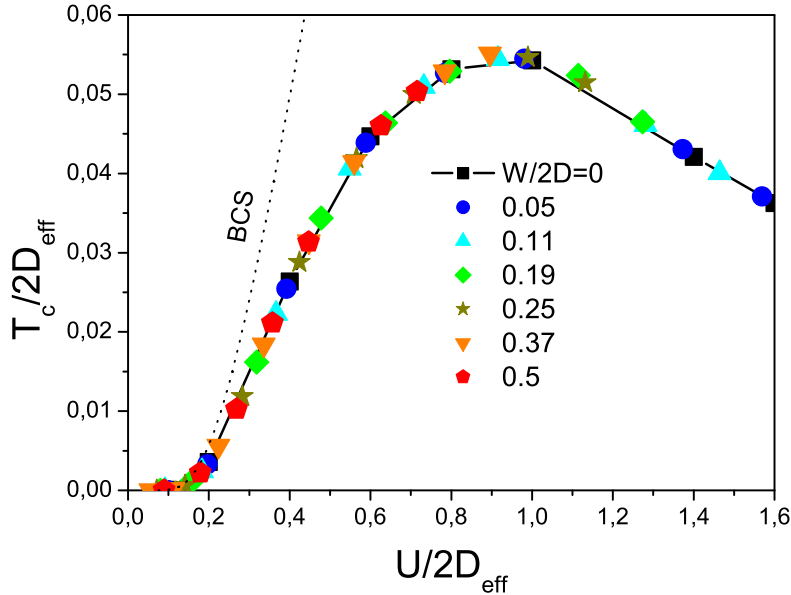}
\caption{Universal dependence of the temperature of superconducting transition
on the strength of Hubbard attraction for different levels of disorder.}
\label{fig1}
\end{figure}

In Ref. \cite{JETP15} it was shown, that disorder influence on the critical
temperature $T_c$ and single -- particle characteristics и (e.g. density of
states) in the model with semi -- elliptic ``bare'' density of states is
universal and reduces only to the change of the effective bandwidth.
In Fig. \ref{fig1}, just for illustrative purposes, we show the universal
dependence of the critical temperature $T_c$ on Hubbard attraction
for different levels of disorder \cite{JETP15}. In the weak coupling region
the temperature of superconducting transition is well described by BCS model
(for comparison in Fig.\ref{fig1} dashed line represent the dependence obtained
for $T_c$ from Eq. (\ref{BCS}) with chemical potential independent of $U$
and determined by quarter filling of the ``bare'' band), while for the strong
coupling region the critical temperature is mainly determined by the condition
of Bose condensation of Cooper pairs and drops with the growth of $U$ as
$t^2/U$, going through the maximum at $U/2D_{eff}\sim 1$.

The review of these and other results obtained for disordered Hubbard model in
DMFT+$\Sigma$ approximation can be found in Ref. \cite{LVK16}.

\section{Ginzburg -- Landau expansion}

Ginzburg -- Landau expansion for the difference of free -- energy densities
of superconducting  and normal states is written in the standard form
\cite{Diagr}:

\begin{equation}
F_{s}-F_{n}=A|\Delta_{\bf q}|^2
+q^2 C|\Delta_{\bf q}|^2+\frac{B}{2}|\Delta_{\bf q}|^4,
\label{GL}
\end{equation}
where $\Delta_{\bf q}$ is the Fourier component of the order parameter.

\begin{figure}
\includegraphics[clip=true,width=\textwidth]{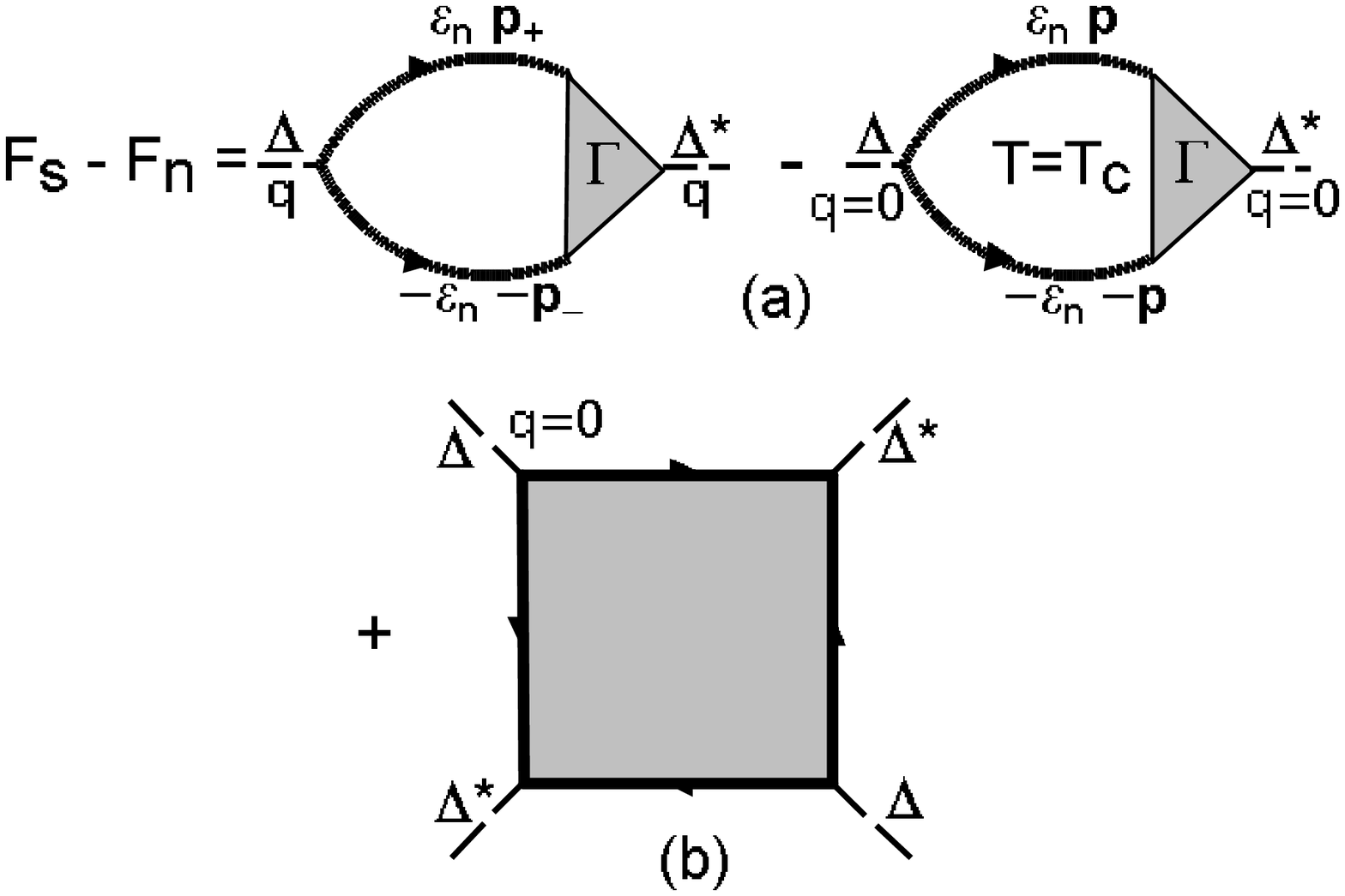}
\caption{Diagrammatic representation of Ginzburg -- Landau expansion.}
\label{diagGL}
\end{figure}

This expansion (\ref{GL}) is determined by by the loop -- expansion diagrams
for free -- energy of an electron in the field of fluctuations of the order --
parameter (denoted by dashed lines) with small wave -- vector ${\bf q}$
\cite{Diagr}, shown in Fig.\ref{diagGL} \cite{Diagr}.
	
In the framework of Nozieres -- Schmitt-Rink approach \cite{NS} we use the weak
coupling approximation to analyze Ginzburg -- Landau coefficients, so that the
``loops'' with two and four Cooper vertices, shown in Fig.\ref{diagGL}, do not
contain contributions from Hubbard attraction and are ``dressed'' only by
impurity scattering. However, like in the case of $T_c$ calculation, the chemical
potential, which is essentially dependent on the coupling strength and in the
strong coupling limit actually controls the condition of Bose condensation of
Cooper pairs, should be determined within full DMFT+$\Sigma$ procedure.

In Ref. \cite{JETP16} it was shown, that in this approach the coefficients
$A$ and $B$ are determined by the following expressions:

\begin{equation}
A(T)=\frac{1}{U}-
\int_{-\infty}^{\infty}d\varepsilon \tilde N_0(\varepsilon)
\frac{th\frac{\varepsilon -\mu }{2T}}{2(\varepsilon -\mu )},
\label{A_end}
\end{equation}

\begin{equation}
B=\int_{-\infty}^{\infty}\frac{d\varepsilon}{2(\varepsilon -\mu)^3}
\left(th\frac{\varepsilon -\mu}{2T}-\frac{(\varepsilon -\mu)/2T}{ch^2
\frac{\varepsilon -\mu}{2T}}\right)
\tilde N_0(\varepsilon).
\label{B_end}
\end{equation}

For $T\to T_c$ the coefficient $A(T)$ takes the usual form:
\begin{equation}
A(T)\equiv \alpha(T-T_c).
\label{A2}
\end{equation} 

In BCS limit, where $T=T_c\to0$, we obtain for coefficients $\alpha$ and $B$ the
standard result \cite{Diagr}:
\begin{equation}
\alpha_{BCS}=\frac{\tilde N_0(\mu)}{T_c}
\qquad B_{BCS}=\frac{7\zeta(3)}{8\pi^2 T_c^2}\tilde N_0(\mu).
\label{aB_BCS}
\end{equation} 

In general case, the coefficients $A$ and $B$ are determined only by the
disorder widened density of states $\tilde N_0(\varepsilon)$ and chemical
potential. Thus, in the case of semi -- elliptic density of states the
dependence of these coefficients on disorder is due only to the simple
replacement $D\to D_{eff}$, leading to universal (independent of the level of
disorder) curves for properly normalized dimensionless coefficients
($\alpha (2D_{eff})^2$ and $B(2D_{eff})^3$) on $U/2D_{eff}$ \cite{JETP16}.
In fact, the coefficients $\alpha$ and $B$ are rapidly suppressed with the
growth of dimensionless coupling $U/2D_{eff}$.

It should be noted. that Eqs. (\ref{A_end}) and (\ref{B_end}) for coefficients
$A$ and $B$ were obtained in Ref. \cite{JETP16} using the exact Ward identities
and remain valid also in the limit of arbitrarily large disorder (including the
region of Anderson localization).

Universal dependence on disorder, related to widening of the band
$D\to D_{eff}$, is observed, in particular, for specific heat discontinuity
at the transition point, which is determined by coefficients $\alpha$ and $B$
\cite{JETP16}:

\begin{equation}
C_s(T_c)-C_n(T_c)=T_c\frac{\alpha^2}{B}.
\label{Cs-Cn}
\end{equation}

\begin{figure}
\includegraphics[clip=true,width=0.8\textwidth]{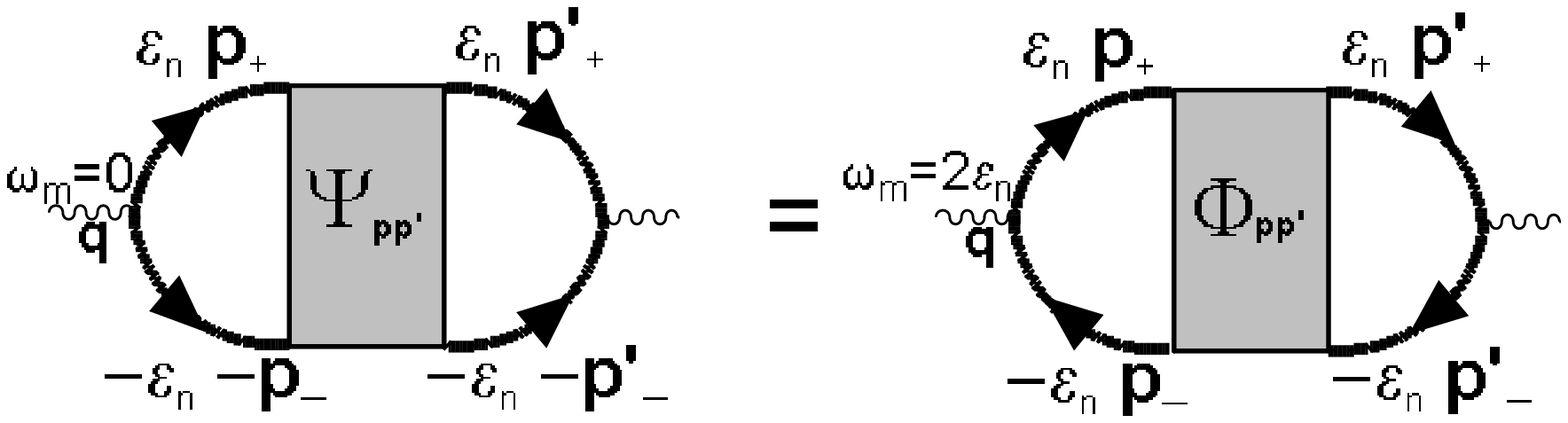}
\caption{The equality of loops in Cooper and diffusion channels under time --
reversal invariance.}
\label{diag_tinv}
\end{figure}

From diagrammatic representation of Ginzburg -- Landau expansion, shown in
Fig.\ref{diagGL} it is clear, that the coefficient $C$ is determined by
the coefficient before $q^2$ in Cooper two -- particle loop
(first term in Fig.\ref{diagGL}). Then we obtain the following expression:

\begin{equation}
C=-T\lim_{q \to 0}\sum_{n, \bf p, \bf {p'}} 
\frac{\Psi_{\bf p\bf {p'}}( \varepsilon_n,{\bf q})-\Psi_{\bf p\bf {p'}}
( \varepsilon_n,0)}{q^2},
\label{C1}
\end{equation}
where $\Psi_{\bf p,\bf {p'}}( \varepsilon_n,{\bf q})$ is two -- particle Green's
function in Cooper channel (see Fig.\ref{diag_tinv}), ``dressed'' in
Nozieres -- Schmitt-Rink approximation only by impurity scattering.
In case of time -- reversal invariance (in the absence of magnetic field and
magnetic impurities) and because of the static nature of impurity scattering
``dressing'' two -- particle Green's function
$\Psi_{\bf p,\bf {p'}}( \varepsilon_n,{\bf q})$, we can reverse here the
direction of all lower electron lines with simultaneous change of the sign of
all momenta (see Fig.\ref{diag_tinv}). As a result we obtain:

\begin{equation}
\Psi_{\bf p,\bf {p'}}( \varepsilon_n,{\bf q})=
\Phi_{\bf p,\bf {p'}}(\omega_m =2\varepsilon_n,{\bf q}),
\label{l1}
\end{equation}
where $\varepsilon_n$ are Fermionic Matsubara frequencies,
${\bf p_{\pm}}={\bf p} \pm \frac{\bf q}{2}$,
$\Phi_{\bf p,\bf {p'}}(\omega_m =2\varepsilon_n,{\bf q})$ is the two --
particle Green's function in diffusion channel, dressed by impurities.
Then we obtain Cooper susceptibility as:

\begin{equation}
\chi({\bf q})=-T\sum_{n, \bf p, \bf {p'}}\Psi_{\bf p,\bf {p'}}
(\varepsilon_n,{\bf q})=-T\sum_{n, \bf p, \bf {p'}}\Phi_{\bf p,\bf {p'}}
(\omega_m =2\varepsilon_n,{\bf q}).
\label{l2}
\end{equation}

Performing the standard summation over Fermionic Matsubara frequencies
\cite{Diagr,AGD}, we obtain:
\begin{equation}
\chi({\bf q})=
-\frac{1}{2\pi}\int_{-\infty}^{\infty}d\varepsilon
Im\Phi^{RA}(\omega =2\varepsilon,{\bf q})th\frac{\varepsilon}{2T},
\label{l3}
\end{equation}
where $\Phi^{RA}(\omega,{\bf q})=\sum_{\bf p, \bf {p'}}
\Phi_{\bf p, \bf {p'}}^{RA}(\omega,{\bf q})$.
To find the loop $\Phi^{RA}(\omega,{\bf q})$ in strongly disordered case
(e.g. in the region of Anderson localization) we can use the approximate
self -- consistent theory of localization \cite{VW,WV,MS,MS86,VW92,Diagr}.
Then this loop contains the diffusion pole of the following form \cite{HubDis}:
\begin{equation}
\Phi^{RA}(\omega=2\varepsilon,{\bf q})=
-\frac{\sum_{\bf {p}}\Delta G_{\bf p}(\varepsilon)}
{\omega +iD(\omega)q^2},
\label{l4}
\end{equation}
where $\Delta G_{\bf p}(\varepsilon)=G^{R}(\varepsilon,{\bf p})-
G^{A}(-\varepsilon,{\bf p})$ and $D(\omega)$ is frequency dependent generalized
diffusion coefficient. Then we obtain the coefficient $C$ as:
\begin{eqnarray}
C=\lim_{q \to 0}\frac{\chi({\bf q})-\chi({\bf q}=0)}{q^2}=
-\frac{1}{8\pi}\int_{-\infty}^{\infty}d\varepsilon
\frac{th\frac{\varepsilon}{2T}}{\varepsilon}Im
\left(
\frac{iD(2\varepsilon)\sum_{\bf {p}}\Delta G_{\bf p}(\varepsilon)}{\varepsilon
+i\delta}
\right)=
\nonumber\\
=-\frac{1}{8\pi}\int_{-\infty}^{\infty}d\varepsilon
\frac{th\frac{\varepsilon}{2T}}{\varepsilon ^2}
Re(D(2\varepsilon)\sum_{\bf {p}}\Delta G_{\bf p}(\varepsilon))
-\frac{1}{16T}Im(D(0)\sum_{\bf {p}}\Delta G_{\bf p}(0)).
\label{l5}
\end{eqnarray}
The generalized diffusion coefficient of self -- consistent theory of
localization \cite{VW,WV,MS,MS86,VW92,Diagr} for our model can be found as the
solution of the following self -- consistency equation \cite{HubDis}:
\begin{equation}
D(\omega)=i\frac{<v>^2}{d}
\left(\omega-\Delta\Sigma_{imp}^{RA}(\omega)+W^4\sum_{\bf p}
(\Delta G_{\bf p}(\varepsilon))^2\sum_{\bf {q}}\frac{1}{\omega+iD(\omega)q^2}
\right)^{-1},
\label{l9}
\end{equation}
where $\omega=2\varepsilon$,
$\Delta\Sigma_{imp}^{RA}(\omega)=\Sigma_{imp}^{R}(\varepsilon)-
\Sigma_{imp}^{A}(-\varepsilon)$, $d$ is space dimension, and velocity
$<v>$ is defined by the following expression:
\begin{equation}
<v>=\frac{\sum_{\bf {p}}|{\bf {v_p}}|\Delta G_{\bf p}(\varepsilon)}
{\sum_{\bf {p}}\Delta G_{\bf p}(\varepsilon)}; 
{\bf {v_p}}=\frac{\partial\varepsilon (\bf p)}{\partial\bf p}.
\label{l7}
\end{equation}
Due to the limits of diffusion approximation summation over $q$ in Eq.
(\ref{l9}) should be limited by the following cut -- off \cite{MS86,Diagr}:
\begin{equation}
q<k_0=Min \{l^{-1},p_F\},
\label{cutoff}
\end{equation}
where $l$ is the mean free path due to elastic disorder scattering and
$p_F$ is Fermi momentum.

In the limit of weak disorder, when localization corrections are small,
the Cooper susceptibility $\chi({\bf q})$ and coefficient $C$ related to it
are determined by the ``ladder'' approximation. In this approximation
coefficient $C$ was studied by us in Ref. \cite{FNT16}, where we obtained
it in  general analytic form. Let us now transform self -- consistency
Eq. (\ref{l9}) to make the obvious connection with exact ``ladder'' expression
in the limit of weak disorder. In ``ladder'' approximation we just neglect
the ``maximally intersecting'' diagrams entering the irreducible vertex
the second term in the r.h.s. of self -- consistency Eq. (\ref{l9}) vanish.
Let us introduce the frequency dependent generalized diffusion coefficient
in ``ladder'' approximation as:
\begin{equation}
D_{0}(\omega)=\frac{<v>^2}{d}
\frac{i}{\omega-\Delta\Sigma_{imp}^{RA}(\omega)}.
\label{l11}
\end{equation}
Then $\frac{<v>^2}{d}$ entering the self -- consistency Eq. (\ref{l9}) can be
rewritten via this diffusion coefficient $D_{0}$ in ``ladder'' approximation,
so that Eq. (\ref{l9}) takes the following form:
\begin{equation}
D(\omega=2\varepsilon)=\frac{D_{0}(\omega=2\varepsilon)}
{1+\frac{W ^4}
{2\varepsilon-\Delta\Sigma_{imp}^{RA}(\omega=2\varepsilon)}
\sum_{\bf p}(\Delta G_{\bf p}(\varepsilon))^{2}
\sum_{\bf q}
\frac{1}{2\varepsilon+iD(\omega=2\varepsilon)q^2}}.
\label{l12}
\end{equation}
Using the approach of Ref. \cite{FNT16} the diffusion coefficient
$D_{0}(\omega=2\varepsilon)$ in ``ladder'' approximation can be derived
analytically. In fact, in ``ladder'' approximation the two -- particle
Green's function (\ref{l4}) takes the following form:
\begin{equation}
\Phi_{0}^{RA}(\omega=2\varepsilon,{\bf q})=
-\frac{\sum_{\bf {p}}\Delta G_{\bf p}(\varepsilon)}
{\omega +iD_{0}(\omega=2\varepsilon)q^2}.
\label{l13}
\end{equation}
Then we obtain:
\begin{equation}
\varphi(\varepsilon,{\bf q}=0)\equiv
\lim_{q \to 0}\frac{\Phi_{0}^{RA}(\omega=2\varepsilon,{\bf q})
-\Phi_{0}^{RA}(\omega=2\varepsilon,{\bf q}=0)}{q^2}=
\frac{i\sum_{\bf {p}}\Delta G_{\bf p}(\varepsilon)}{{\omega}^2}D_{0}
(\omega=2\varepsilon).
\label{l14}
\end{equation}
Then the diffusion coefficient $D_{0}$ can be written as:
\begin{equation}
D_{0}=\frac{\varphi(\varepsilon,{\bf q}=0)(2\varepsilon)^2}
{i\sum_{\bf {p}}\Delta G_{\bf p}(\varepsilon)}.
\label{l15}
\end{equation}
In Ref. \cite{FNT16} using the exact Ward identity we have shown, that in
``ladder'' approximation $\varphi(\varepsilon,{\bf q}=0)$ can be represented as:
\begin{equation}
\varphi(\varepsilon,{\bf q}=0)(2\varepsilon)^2=\sum_{\bf {p}}v_{x}^2
G^{R}(\varepsilon,{\bf p})G^{A}(-\varepsilon,{\bf p})+
\frac{1}{2}\sum_{\bf {p}}
\frac{\partial^2\varepsilon (\bf p)}{\partial p_{x}^2}
(G^{R}(\varepsilon,{\bf p})+G^{A}(-\varepsilon,{\bf p})),
\label{l16}
\end{equation}
where $v_x=\frac{\partial\varepsilon (\bf p)}{\partial p_{x}}$.

Finally, using Eqs. (\ref{l16}), (\ref{l15}) we find the diffusion coefficient
$D_{0}$ in  ``ladder'' approximation. Using self -- consistency Eq. (\ref{l12})
we determine the generalized diffusion coefficient, and then using
Eq. (\ref{l5}) we find the coefficient $C$. In the limit of weak disorder,
when ``ladder'' approximation works well and generalized diffusion coefficient
just coincides with diffusion coefficient in ``ladder'' approximation, we
obtain for coefficient $C$ the result obtained in Ref. \cite{FNT16}:
\begin{eqnarray}
C_{0}=-\frac{1}{8\pi}
\int_{-\infty}^{\infty}d\varepsilon\frac{th\frac{\varepsilon}{2T}}{\varepsilon^2}
\sum_{\bf p}
\left(
{v_x}^2Im(G^{R}(\varepsilon,{\bf p})G^{A}(-\varepsilon,{\bf p}))
+\frac{{\partial}^2\varepsilon_{\bf p}}{\partial {p_x}^2}
ImG^{R}(\varepsilon,{\bf p})
\right)+
\nonumber\\
\frac{1}{16T}\sum_{\bf p}
\left(
{v_x}^2Re(G^{R}(0,{\bf p})G^{A}(0,{\bf p}))+
\frac{{\partial}^2\varepsilon_{\bf p}}{\partial {p_x}^2}ReG^{R}(0,{\bf p})
\right).
\label{l17}
\end{eqnarray}
Now we can use the iteration scheme to find the coefficient $C$, which in the
limit of weak disorder reproduce the results ``ladder'' approximation, while
in the limit of strong disorder takes into account the effects of Anderson
localization (in the framework of self -- consistent theory of localization).

In numerical calculations using Eqs. (\ref{l15}) and (\ref{l16}) we first find
the ``ladder'' diffusion coefficient $D_{0}$ for the given value of
$\omega=2\varepsilon$. Then, solving by iterations the transcendental
self -- consistency Eq. (\ref{l12}), we determine the generalized diffusion
coefficient at this frequency. After that, using Eq. (\ref{l5}) we calculate
Ginzburg -- Landau coefficient $C$.

In Ref. \cite{HubDis} it was shown, that in DMFT+$\Sigma$ approximation for
Anderson -- Hubbard model the critical disorder for Anderson metal -- insulator
transition $W/2D=0.37$ and is independent of the value of Hubbard interaction
$U$. The approach developed here allows determination of $C$ coefficient
also in the region of Anderson insulator at disorder levels $W/2D>0.37$.

\section{Main results}

The coherence length at given temperature $\xi(T)$ gives a characteristic scale
of inhomogeneities of the order parameter $\Delta$:
\begin{equation}
\xi^2(T)=-\frac{C}{A}.
\label{xi2}
\end{equation}
Coefficient $A$ changes its sign and becomes zero at critical temperature:
$A=\alpha(T-T_c)$, so that
\begin{equation}
\xi(T)=\frac{\xi}{\sqrt{1-T/T_c}},
\label{xiT}
\end{equation}
where we have introduced the coherence length of a superconductor:
\begin{equation}
\xi=\sqrt{\frac{C}{\alpha T_c}},
\label{xi0}
\end{equation}
which reduces to a standard expression in the weak coupling region and in the
absence of disorder \cite{Diagr}:
\begin{equation}
\xi_{BCS}=\sqrt{\frac{C_{BCS}}{\alpha_{BCS} T_c}}=\sqrt{\frac{7\zeta(3)}
{16\pi^2 d}}\frac{v_F}{T_c}.
\label{xi_BCS}
\end{equation}

Penetration depth of magnetic field into superconductor is defined by:

\begin{equation}
\lambda^2(T)=-\frac{c^2}{32 \pi e^2}\frac{B}{A C}.
\label{lambda2}
\end{equation}
Then:
\begin{equation}
\lambda (T)=\frac{\lambda}{\sqrt{1-T/T_c}},
\label{lambdaT}
\end{equation}
where we have introduced:
\begin{equation}
\lambda^2=\frac{c^2}{32 \pi e^2}\frac{B}{\alpha C T_c},
\label{lambda0}
\end{equation}
which in the absence of disorder has the form:
\begin{equation}
\lambda^2_{BCS}=\frac{c^2}{32 \pi e^2}\frac{B_{BCS}}{\alpha_{BCS}
C_{BCS} T_c}=\frac{c^2}{16 \pi e^2}\frac{d}{N_0(\mu)v_F^2}.
\label{lambda_BCS}
\end{equation}
As $\lambda_{BCS}$ is independent of $T_c$, i.e. of coupling strength, it is
convenient to use for normalization of penetration depth
$\lambda$ (\ref{lambda0}) at arbitrary $U$ and $W$.

Close to $T_c$ the upper critical magnetic field $H_{c2}$ is determined by
Ginzburg -- Landau coefficients as:
\begin{equation}
H_{c2}=\frac{\Phi_0}{2 \pi \xi^2(T)}=-\frac{\Phi_0}{2 \pi}\frac{A}{C},
\label{Hc2}
\end{equation}
where $\Phi_0=c \pi/e$ is magnetic flux quantum.
Then the slope of the upper critical filed close to $T_c$ is given by:
\begin{equation}
\frac{dH_{c2}}{dT}= \frac{\Phi_0}{2 \pi}\frac{\alpha}{C}.
\label{dHc2}
\end{equation}

\begin{figure}
\includegraphics[clip=true,width=0.7\textwidth]{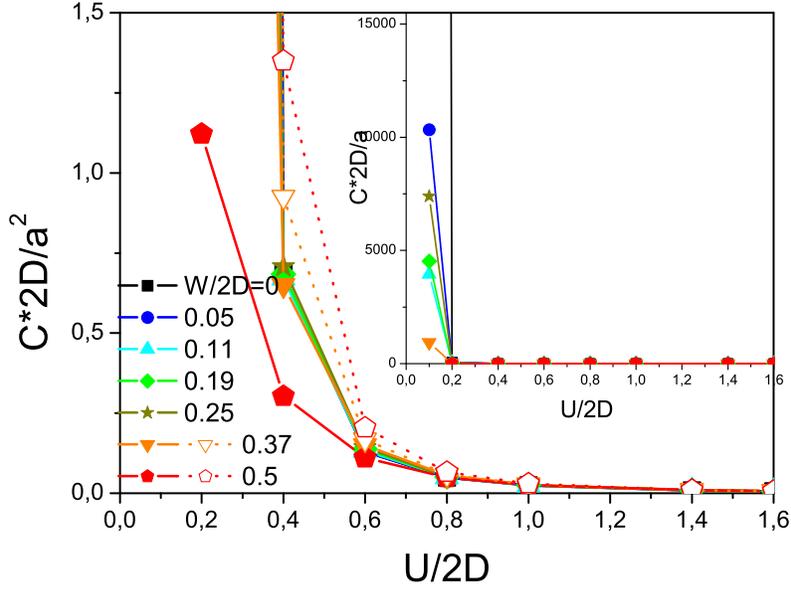}
\caption{Dependence of $C$ coefficient on the strength of Hubbard attraction
for different levels of disorder ($a$ is lattice parameter). Filled symbols
and continuous lines correspond to calculations taking into account
localization corrections. Unfilled symbols and dashed lines correspond to
``ladder'' approximation.}
\label{fig5}
\end{figure}

In Fig.\ref{fig5} we show the dependence of coefficient $C$ on the strength of
Hubbard attraction for different disorder levels. On this figure and in the
following we use filled symbols and continuous lines correspond to the results of
calculations taking into account localization corrections, while unfilled
symbols and dashed lines correspond to calculations in ``ladder'' approximation.
Coefficient $C$ is essentially two -- particle characteristic and it does not
follow universal behavior on disorder, as in case of coefficients $A$ and $B$,
and disorder dependence here is not reduced only to widening of effective
bandwidth by disorder. Correspondingly, the dependence of $C$ on coupling
strength, where all energies are normalized by effective bandwidth $2D_{eff}$,
we do not observe a universal curve for different levels of disorder \cite{FNT16},
in contrast to similar dependencies for coefficients $\alpha$ and $B$. In fact,
coefficient $C$ is rapidly suppressed with the growth of coupling strength.
Especially strong suppression is observed in weak coupling region
(cf. insert in Fig.\ref{fig5}). Localization corrections become relevant in the
limit of strong enough disorder ($W/2D>0.25$). Under such strong disordering
localization corrections significantly suppress coefficient $C$ in weak
coupling region (cf. dashed lines (``ladder'' approximation) and continuous
curves (with localization corrections) for $W/2D=0.37$ and $0.5$)
In strong coupling region for $U/2D>1$ localization corrections, in fact, do
not change the value of coefficient $C$, as compared to the results of
``ladder'' approximation, even in the limit of strong disorder for $W/2D>0.37$,
where the system becomes Anderson insulator. 

\begin{figure}
\includegraphics[clip=true,width=0.7\textwidth]{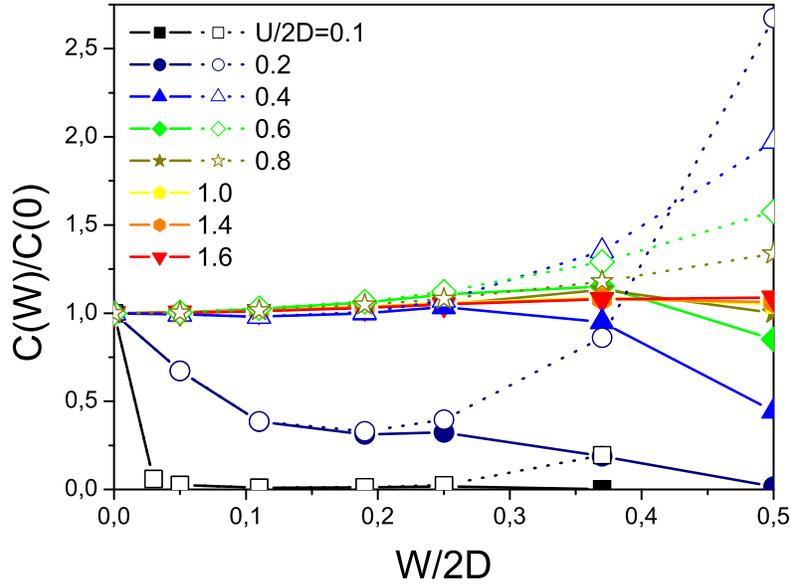}
\caption{Dependence of coefficient $C$ normalized by its value in the absence of
disorder for different values of Hubbard attraction $U$.
Dashed lines -- ``ladder'' approximation, continuous curves -- calculations with
the account of localization corrections.}
\label{fig6}
\end{figure}
In Fig.\ref{fig6} we show the dependencies of coefficient $C$ on disorder level
for different values of coupling strength $U/2D$. In the limit of weak coupling
($U/2D=0.1$) we observe rather rapid suppression of coefficient $C$ with the
growth of disorder in case of weak enough impurity scattering.  In the
region of strong enough disorder in ``ladder'' approximation we can observe
some growth of coefficient $C$ with the increase of disorder, which is related
mainly with significant widening of the band by such strong disorder and
corresponding drop of the effective coupling $U/2D_{eff}$.
However, localization corrections, which are significant at large disorder
$W/2D>0.25$, actually lead to suppression of coefficient $C$ with the growth of
disorder in the limit of strong impurity scattering. In the intermediate
coupling region ($U/2D=0.4 - 0.6$) coefficient $C$ in ``ladder'' approximation
is only slightly growing with increasing disorder. In BEC limit ($U/2D>1$)
coefficient $C$ is practically independent of impurity scattering both in
``ladder'' approximation and with the account of localization corrections.
In BEC limit the account of localization corrections in fact do not change the
value of $C$ in comparison with ``ladder'' approximation.

As Ginzburg -- Landau expansion coefficient $\alpha$ and $B$ demonstrate the
universal dependence on disorder, Anderson localization in fact does not
influence them at all, while coefficient $C$ in the weak coupling region is
strongly affected by localization corrections, being almost independent of them
in BEC limit, the physical properties depending on $C$ will be also
significantly changed by localization corrections in the weak coupling region,
becoming practically independent of localization in BEC limit.

\begin{figure}
\includegraphics[clip=true,width=0.7\textwidth]{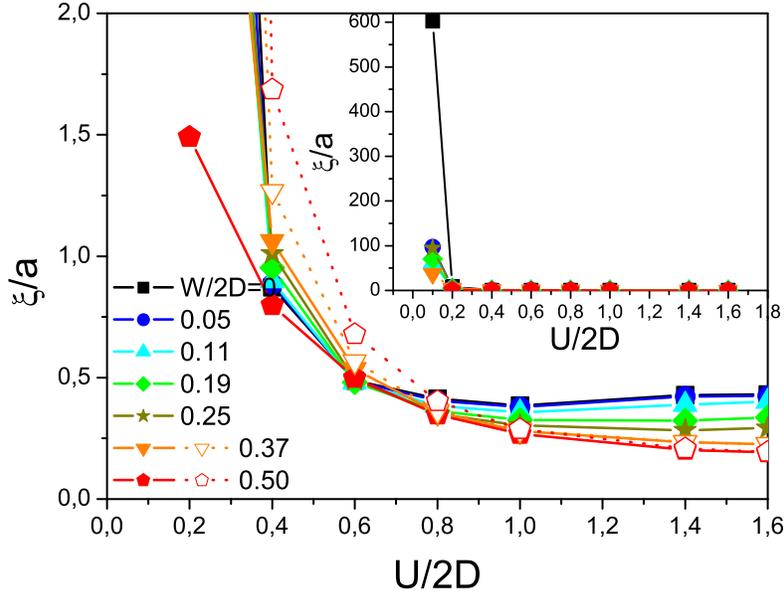}
\caption{Dependence of coherence length on the strength of Hubbard attraction
$U$ for different disorder levels. Insert: the rapid growth of coherence length
with diminishing coupling in BCS limit.}
\label{fig7}
\end{figure}

Let us now discuss the behavior of physical properties. Dependence of coherence
length on Hubbard attraction strength is shown in Fig.\ref{fig7}. We can see
that in the weak coupling region (cf. insert at Fig.\ref{fig7}) coherence
length rapidly drops with the growth of $U$ for any disorder, reaching the
value of the order of lattice parameter $a$ in the intermediate coupling
region of $U/2D \sim 0.4-0.6$. Further growth of coupling strength changes
the coherence length only slightly. The account of localization corrections
for coherence length is significant only at large disorder ($W/2D>0.25$).
We see, that localization corrections lead to significant suppression of
coherence length in BCS limit of weak coupling and practically do not change
the coherence length in BEC limit.

\begin{figure}
\includegraphics[clip=true,width=0.7\textwidth]{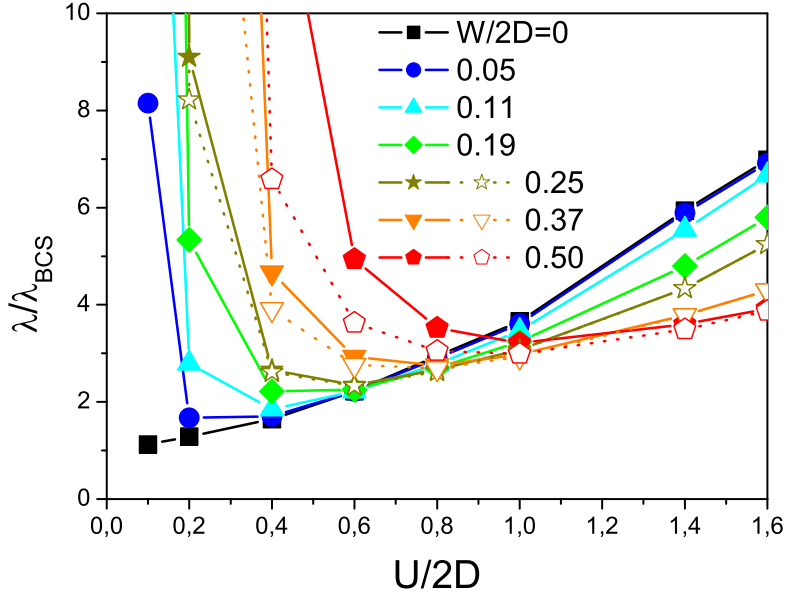}
\caption{Dependence of penetration depth, normalized by its BCS value in the
limit of weak coupling, on the strength of Hubbard attraction $U$ for
different levels of disorder.}
\label{fig8}
\end{figure}

In Fig.\ref{fig8} we show the dependence of penetration depth, normalized by its
BCS value in the absence of disorder (\ref{lambda_BCS}), on the strength of
Hubbard attraction $U$ for different levels of disorder. In the absence of
impurity scattering penetration depth grows with the increase of the coupling
strength. In BCS weak coupling limit disorder leads to a fast growth of
penetration depth (for ``dirty'' BCS superconductors $\lambda\sim l^{-1/2}$,
where $l$ is the mean free path). In BEC strong coupling limit disorder only
slightly diminish the penetration depth (cf. Fig.\ref{fig11}(a)).
This leads to suppression of penetration depth with disorder with the growth of
Hubbard attraction strength in the region of weak enough coupling and to the
growth of $\lambda$ with $U$ in BEC strong coupling region. The account of
localization corrections is significant only in the limit of strong disorder
($W/2D>0.25$) and leads to noticeable growth of penetration depth as
compared to the ``ladder'' approximation in the weak coupling region. In BEC
limit the influence of localization on penetration depth is just insignificant.

\begin{figure}
\includegraphics[clip=true,width=0.7\textwidth]{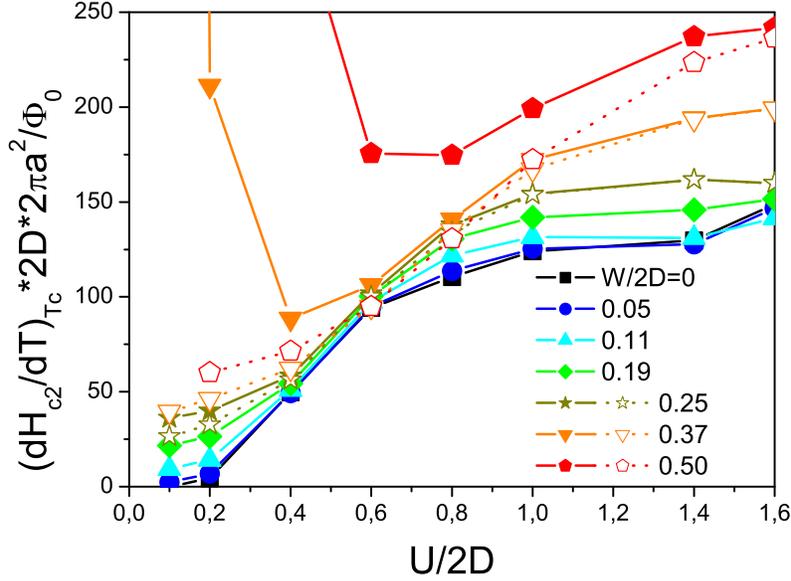}
\caption{Dependence of the slope of the upper critical field on the strength
of Hubbard attraction $U$ for different level of disorder.}
\label{fig9}
\end{figure}

Dependence of the slope of the upper critical magnetic field on the strength of
Hubbard attraction for different disorder levels is shown in Fig.\ref{fig9}.
In the limit of weak enough impurity scattering, until Anderson localization
corrections remain unimportant, the slope of the upper critical field grows
with the growth of the coupling strength. The fast growth of the slope is
observed with the growth of $U$ in the region of weak enough coupling, while
in the limit of strong coupling the slope is rather weakly dependent on $U/2D$.
In the region of strong enough disorder ($W/2D>0.25$) the account of
localization corrections becomes quite important -- it qualitatively changes
the behavior of the upper critical field. While ``ladder'' approximation
(dashed curves) conserves the behavior of the slope of the upper critical field
typical for the region of weak disorder, where the slope grows with the
growth of the coupling strength, the account of Anderson localization
($W/2D \geq 0.37$) leads to the strong increase of the slope of the upper
critical field in the weak coupling limit. As a result, in Anderson insulator
the slope of the upper critical filed rapidly drops with the growth of $U$
in the weak coupling limit and just insignificantly grows with the growth of
$U$ in BEC limit. Note that the account of localization corrections is
also unimportant for for the slope of the upper critical field in the strong
coupling limit.

\begin{figure}
\includegraphics[clip=true,width=0.48\textwidth]{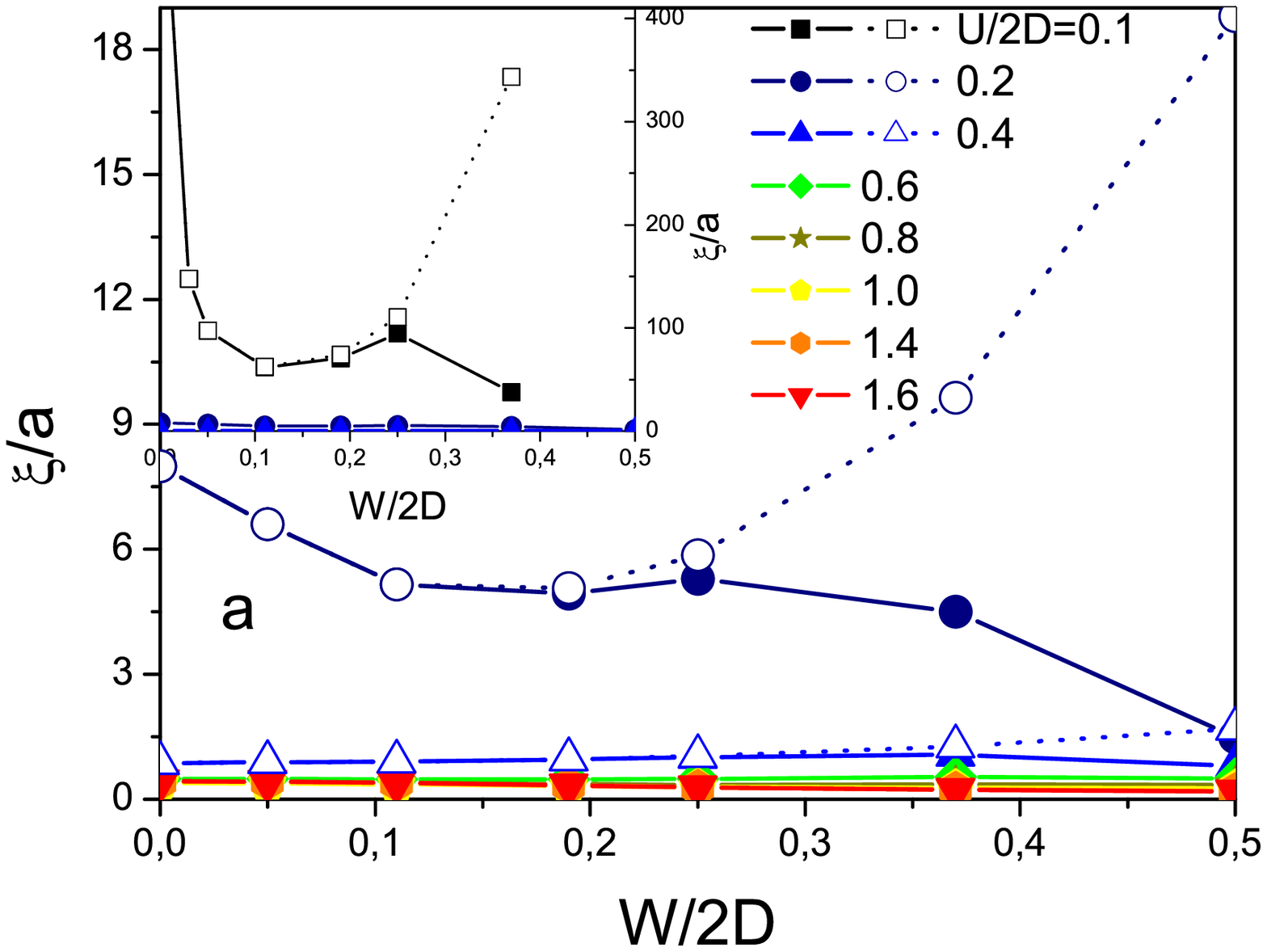}
\includegraphics[clip=true,width=0.48\textwidth]{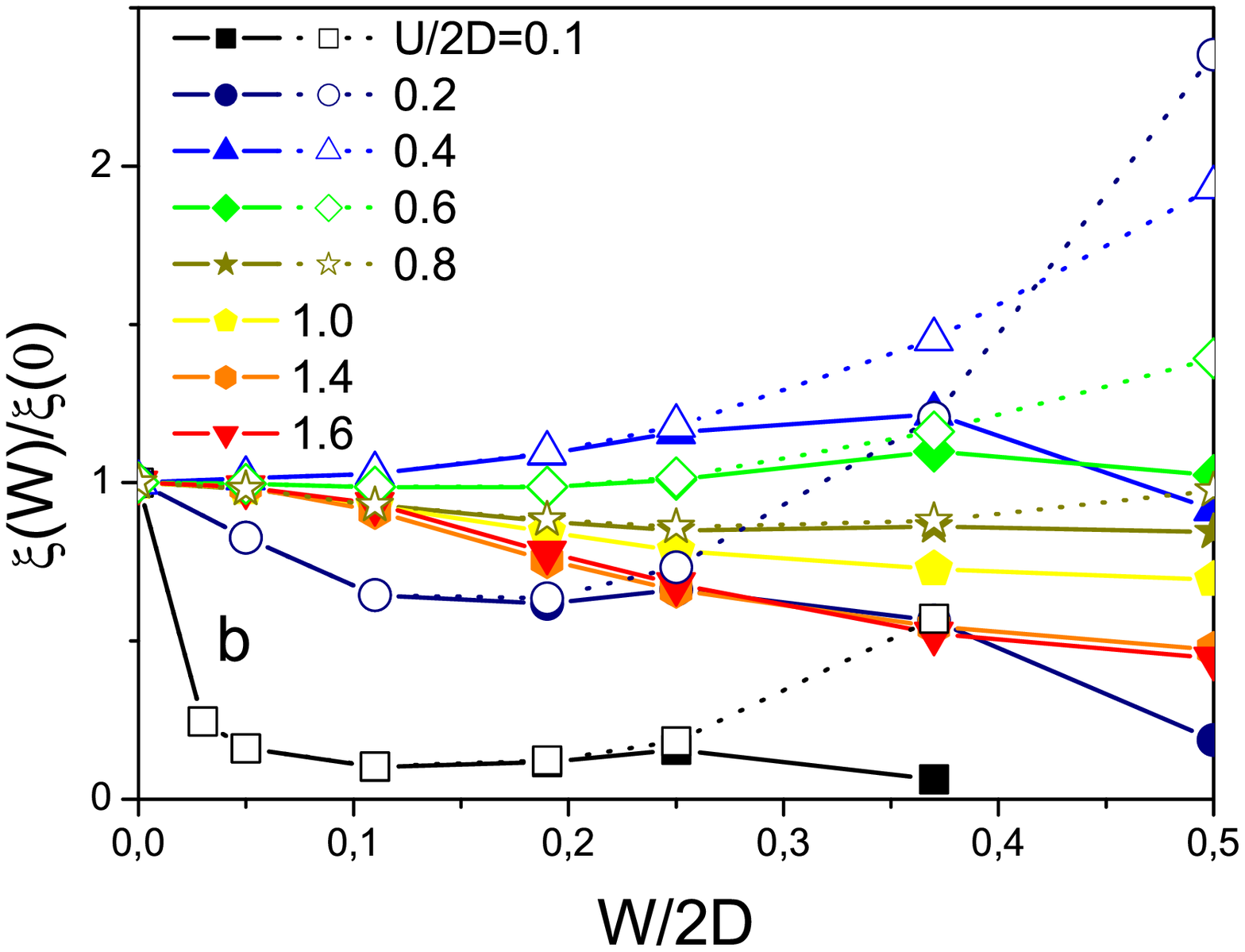}
\caption{Dependence of coherence length on disorder for different values of
Hubbard attraction.
(a) --- coherence length normalized by lattice parameter $a$.
Insert: dependence of coherence length on disorder in weak coupling limit.
(b) --- coherence length normalized by its value in the absence of disorder.}
\label{fig10}
\end{figure}

Let us consider now dependencies of physical properties on disorder.
In Fig.\ref{fig10} we show dependence of coherence length $\xi$ on disorder
for different values of coupling. In BCS limit for weak coupling and for
weak enough impurity scattering we observe the standard ``dirty'' superconductor
dependence $\xi \sim l^{1/2}$, i.e. coherence length rapidly drops with the
growth of disorder (cf. insert in Fig.\ref{fig10}(a)). However, at strong
enough disorder in ``ladder'' approximation (dashed lines) coherence length
starts to grow with disorder (cf. insert in Fig.\ref{fig10}(a) and
Fig.\ref{fig10}(b)), which is mainly related to the widening of the band by
disorder and corresponding suppression of $U/2D_{eff}$.
Taking into account localization corrections leads to noticeable suppression of
coherence length in comparison with ``ladder'' approximation in the limit of
strong disorder, which leads to restoration of general suppression of $\xi$
with the growth of disorder in this limit. In standard BCS model with bare band
of infinite width coherence length drops with the growth of disorder
$\xi \sim l^{1/2}$ and close to Anderson transition this suppression of $\xi$
even accelerates, so that $\xi \sim l^{2/3}$ \cite{SCLoc_1,SCLoc_2,SCLoc_3},
which differs from the present model here, where close to Anderson coherence
length is rather weakly dependent on disorder, which is related to significant
widening of the band by disorder. With growth of coupling, for $U/2D \geq0.4-0.6$
coherence length $\xi$ becomes of the order of lattice parameter and is almost
disorder independent, while in BEC limit of very strong coupling $U/2D=1.4, 1.6$
the growth of disorder up to very strong values ($W/2D=0.5$) leads to suppression
of coherence length approximately by the factor of two (cf. Fig.\ref{fig10}(b)).
Again we see, that in the limit of strong coupling the account of localization
corrections is rather insignificant.

\begin{figure}
\includegraphics[clip=true,width=0.48\textwidth]{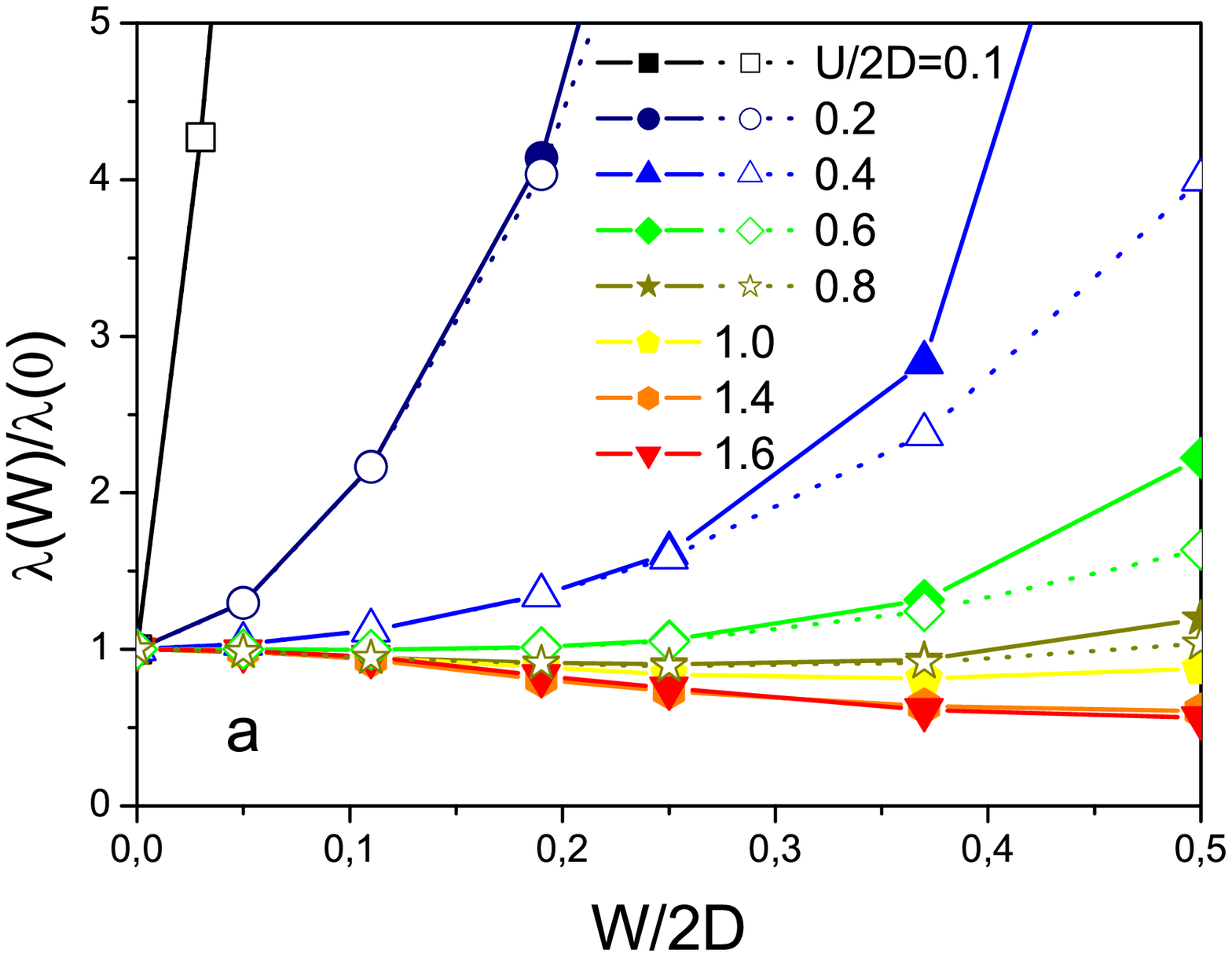}
\includegraphics[clip=true,width=0.48\textwidth]{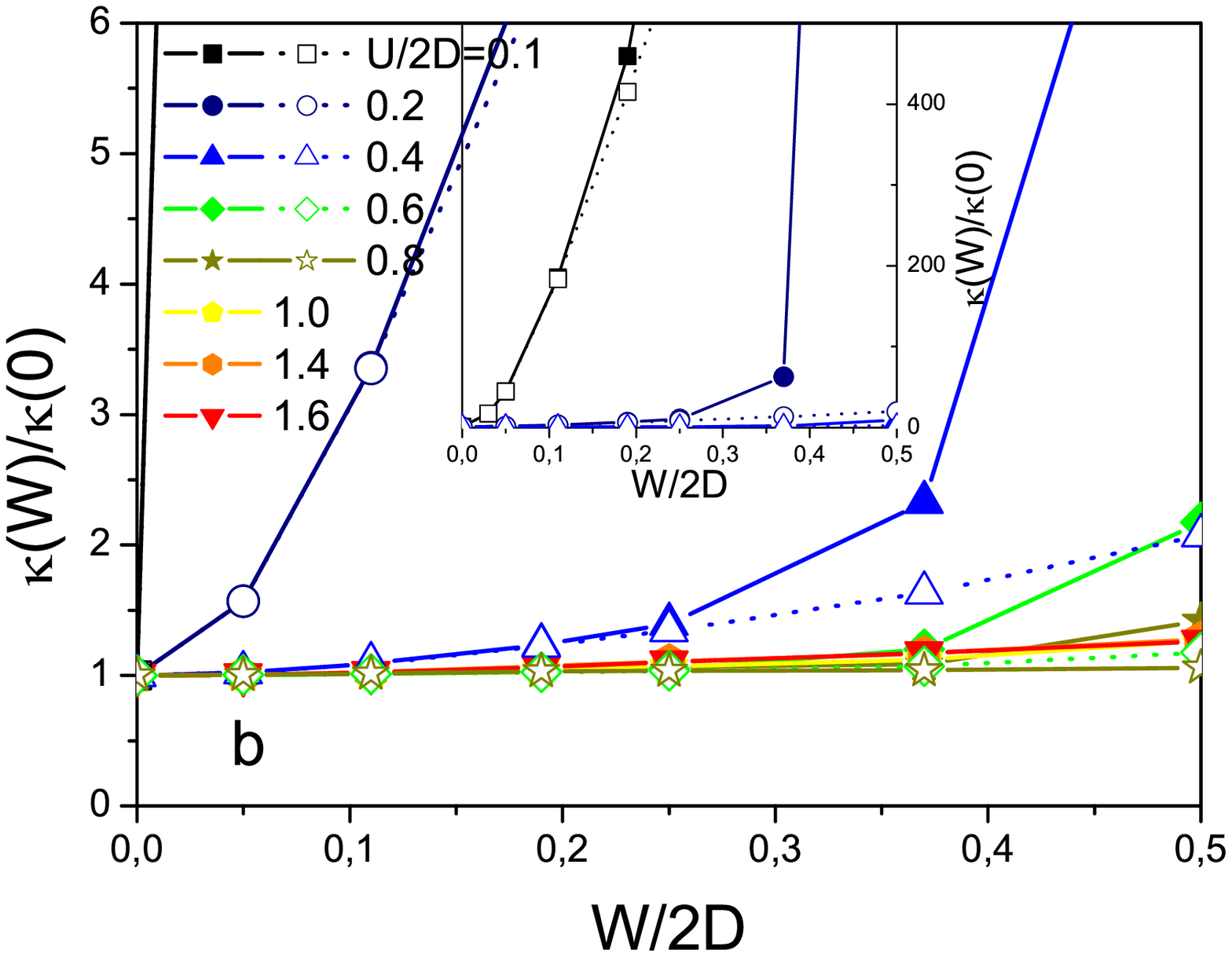}
\caption{Dependence of penetration depth (a) and Ginzburg -- Landau parameter
(b) on disorder level for different values of Hubbard attraction.
 Insert shows the growth of Ginzburg -- Landau parameter with disorder in weak
 coupling limit.}
\label{fig11}
\end{figure}

Dependence of penetration depth on disorder for different values of Hubbard
attraction is shown in Fig.\ref{fig11}(a). In weak coupling limit disorder
in accordance with the theory of ``dirty'' superconductors leads to the growth
of penetration depth ($\lambda \sim l^{-1/2}$). With increase of the coupling
strength the growth of penetration depth slow down and in the limit of very
strong coupling, for $U/2D=1.4, 1.6$, penetration depth is even slightly
suppressed by disorder. The account of localization corrections leads to some
quantitative growth of penetration depth in comparison with the results of
``ladder'' approximation in the weak coupling region. Qualitatively the
dependence of penetration depth on disorder does not change. In BEC limit of
strong coupling the account of localization corrections is rather irrelevant.
In Fig.\ref{fig11}(b) we show the disorder dependence of dimensionless
Ginzburg -- Landau $\kappa = \lambda / \xi$. We can see, that in the weak
coupling limit Ginzburg -- Landau parameter is rapidly growing with disorder
(cf. insert in Fig.\ref{fig11}(b)) in accordance with the theory of ``dirty''
superconductors, where $\kappa \sim l^{-1}$. With the increase of coupling
strength the growth of Ginzburg -- Landau parameter with disorder slows down and
in the limit of strong coupling $U/2D>1$ parameter $\kappa$ is practically
disorder independent. The account of localization corrections quantitatively
increases Ginzburg -- Landau parameter in Anderson insulator phase
($W/2D \geq 0.37$) in the strong coupling region. In the strong coupling region
localization corrections are again irrelevant.

\begin{figure}
\includegraphics[clip=true,width=0.48\textwidth]{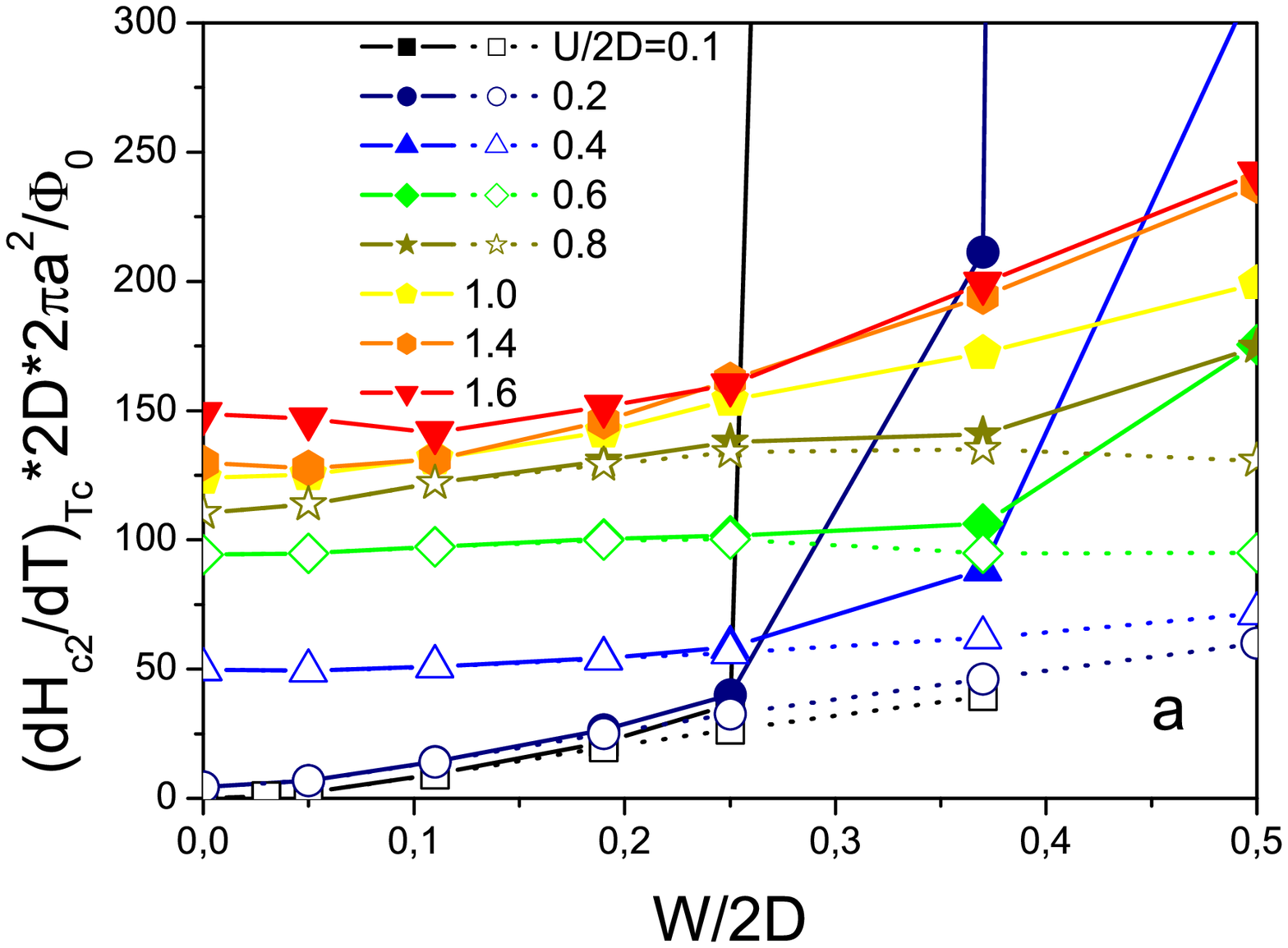}
\includegraphics[clip=true,width=0.48\textwidth]{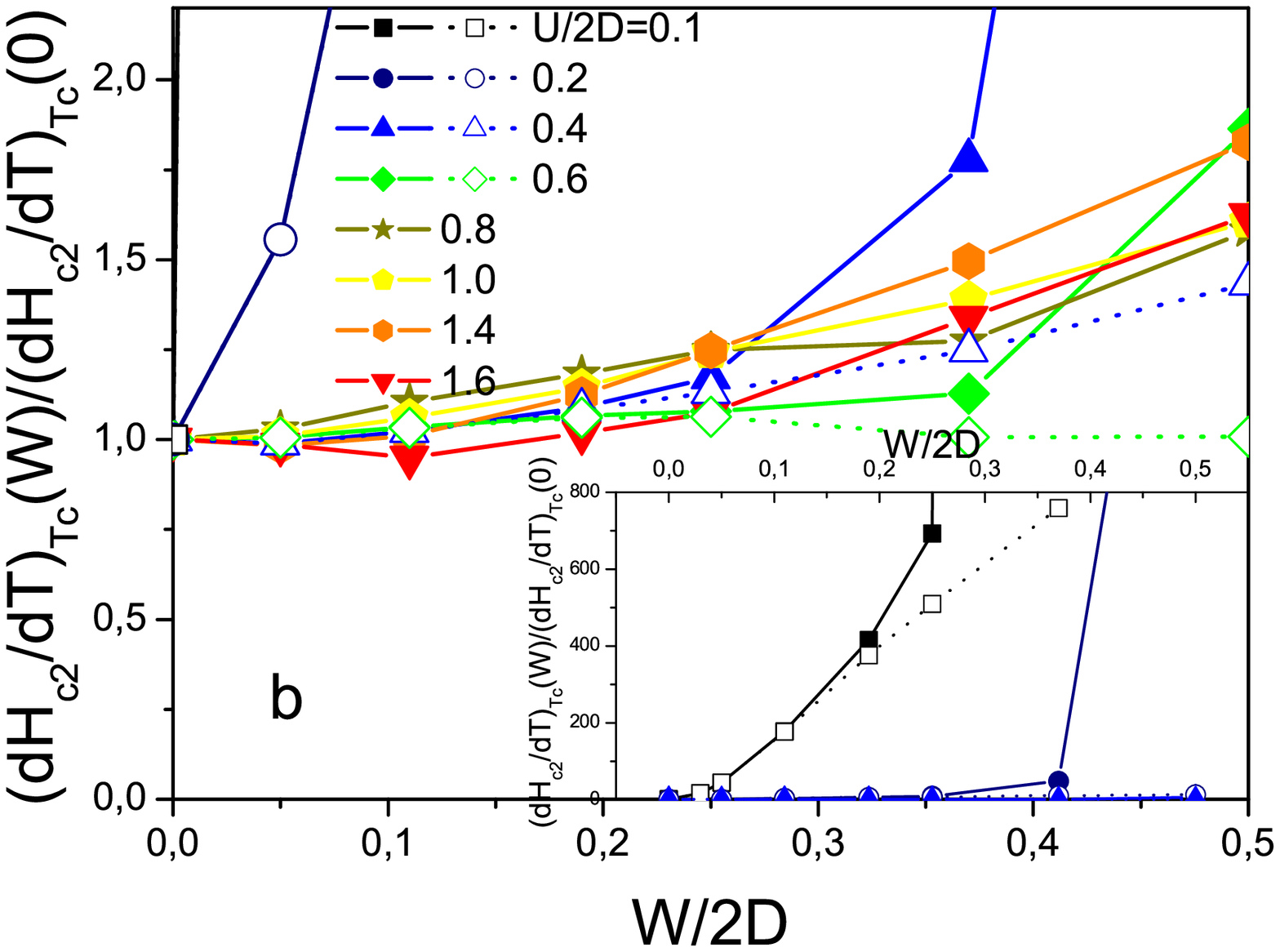}
\caption{Dependence of the slope of the upper critical field (a) and this
slope, normalized by its value in the absence of disorder (b), on disorder
for different values of Hubbard attraction strength.
In the insert we show the growth of the slope with disorder in weak coupling
region.}
\label{fig12}
\end{figure}

In Fig.\ref{fig12} we show the disorder dependence of the slope of the upper
critical field. In the weak coupling limit we again observe the behavior typical
for ``dirty'' superconductors --- the slope of the upper critical field grows
with the growth of disorder (cf. Fig.\ref{fig12}(a) and the insert in
Fig.\ref{fig12}(b)). The account of localization corrections in weak coupling
limit sharply increases the slope of the upper critical field in comparison with
the result of ``ladder'' approximation in the region of Anderson insulator
($W/2D \geq 0.37$). As a result, in Anderson insulator the slope of the upper
critical field grows with the increase of impurity scattering much faster,
than in ``ladder'' approximation. In intermediate coupling region
($U/2D=0.4 - 0.8$) the slope of the upper critical field is practically
independent of impurity scattering in the region of weak disorder.
In ``ladder'' approximation such behavior is conserved also in the region of
strong disorder. However, the account of localization corrections leads to
significant growth of the slope with disorder in Anderson insulator phase.
In the limit of very strong coupling and weak disorder the slope of the upper
critical field can even slightly diminish with disorder, but in the limit of
strong disorder the slope grows with growth of impurity scattering. In BEC
limit the account of localization corrections is irrelevant and only slightly
changes the slope of the upper critical field as compared with the results of
``ladder'' approximation.

\section{Conclusion}

In this paper in the framework of Nozieres -- Schmitt-Rink approximation and
DMFT+$\Sigma$ generalization of dynamical mean field theory we have studied the
effects of disorder (including the strong disorder region of Anderson
localization) on Ginzburg -- Landau coefficients and related physical properties
close to $T_c$ in disordered Anderson -- Hubbard model with attraction.
Calculations were done for the wide range of attractive potentials $U$,
from weak coupling region $U/2D_{eff}\ll 1$, where instability of normal phase
and superconductivity is well described by BCS model, up to the strong coupling
limit $U/2D_{eff}\gg 1$, where transition into superconducting state is due to
Bose condensation of compact Cooper pairs, forming at temperature much higher
than the temperature of superconducting transition.

The growth of the coupling strength $U$ leads to rapid suppression of all
Ginzburg -- Landau coefficients. The coherence length $\xi$ rapidly drops with
the growth of coupling and for $U/2D \sim 0.4$ becomes of the order of lattice
spacing and only slightly changes with further increase of coupling.
Penetration depth in ``clean'' superconductors grows with $U$, while in
``dirty'' superconductors it drops in the weak coupling and grows in BEC limit,
passing through the minimum in the intermediate coupling region
$U/2D\sim 0.4-0.8$. In the region of weak enough disorder ($W/2D<0.37$), when
Anderson localization effect are not much important, the slope of the upper
critical field grows with the growth of $U$. However, in the limit of weak
coupling in Anderson insulator phase localization effects sharply increase the
slope of the upper critical field, while in BEC limit of strong coupling
localization effects become unimportant. As a result, the slope of the upper
critical field drops with the growth of $U$ in BCS limit, passing through
the minimum at $U/2D\sim 0.4-0.8$. The specific heat discontinuity grows with
Hubbard attraction $U$ in the weak coupling region and drops in the strong
coupling limit, passing through the maximum at $U/2D_{eff}\approx 0.55$
\cite{JETP16}.

Disorder influence (including the strong disorder in the region of Anderson
localization) upon the critical temperature $T_c$ and Ginzburg -- Landau
coefficients $A$ and $B$ and the related discontinuity of specific heat is
universal and is completely determined only by disorder widening
of the bare band, i.e. by the replacement $D \to D_{eff}$. Thus, even in the
strong coupling region, the critical temperature and Ginzburg -- Landau
coefficients $A$ and $B$ satisfy the generalized Anderson theorem --- all
influence of disorder is related only to the change of the density of states.
Disorder influence on coefficient $C$ is not universal and is related not only
to the bare band widening.

Coefficient $C$ is sensitive to the effects of Anderson localization.
We have studied this effect in for a wide range of disorder, including the
region of Anderson insulator. To compare and extract explicitly effects of
Anderson localization we also studied coefficient $C$ in ``ladder''
approximation for disorder scattering. In the weak coupling limit
$U/2D_{eff}\ll 1$ and weak disorder $W/2D<0.37$ the behavior of coefficient $C$
and related physical properties is well described by the theory of ``dirty''
superconductors -- coefficient $C$ and coherence length rapidly drop with the
growth of disorder, while penetration depth and the slope of the upper critical
field grow. In the region of strong disorder (in Anderson insulator) in BCS
limit the behavior of coefficient $C$ is strongly affected by localization
effects. In ``ladder'' approximation the band widening effect leads to the
growth of coefficient $C$ with the growth of $W$ \cite{FNT16}, however
localization effects restore suppression of coefficient $C$ by disorder and
in Anderson insulator phase. Correspondingly, localization effects significantly
change physical properties, related to coefficient $C$, so that for these
properties qualitatively follow the dependencies characteristic for ``dirty''
superconductors --- the coherence length is suppressed by disorder, while
the penetration depth and the slope of the upper critical field grow with the
growth of disorder. In BCS -- BEC crossover region and in BEC limit coefficient
$C$ and all related physical properties are rather weakly dependent on disorder.
In particular, in BEC limit both coherence length and penetration depth are
slightly suppressed by disorder, so that their ratio (Ginzburg -- Landau
parameter) is practically disorder independent. In BEC limit the effects of
Anderson localization rather weakly affect the coefficient $C$ and the related
physical characteristics.

It should be noted, that all results were derived here under implicit
assumption of self -- averaging nature of superconducting order parameter
entering Ginzburg -- Landau expansion, which is connected with our use of the
standard ``impurity'' diagram technique \cite{Diagr,AGD}. It is well known
\cite{SCLoc_3}, that this assumption becomes, in general case, inapplicable
close to Anderson metal -- insulator transition, due to strong
fluctuations of the local density of states developing here \cite{NAV_1}
and inhomogeneous picture of superconducting transition \cite{NAV_2}.
This problem is very interesting in the context of the superconductivity
in BCS -- BEC crossover region and in the region of strong coupling and
deserves further studies.

This work was supported by RSF grant 14-12-00502.

\newpage

\end{document}